\documentclass[12pt]{article}

\usepackage{graphicx}
\begin{document}
\begin{center}
{\bf Nonlinear electrodynamics with birefringence}\\
\vspace{5mm}
 S. I. Kruglov
\footnote{E-mail: serguei.krouglov@utoronto.ca}
 \\

\vspace{5mm}
\textit{Department of Chemical and Physical Sciences, University of Toronto,\\
3359 Mississauga Road North, Mississauga, Ontario, Canada L5L 1C6}
\end{center}

\begin{abstract}

A new model of nonlinear electrodynamics with three parameters is
suggested. The phenomena of vacuum birefringence takes place when there is the external constant magnetic field.
We calculate the indices of refraction for two polarizations of electromagnetic waves, parallel and perpendicular
to the magnetic induction field. From the Bir\'{e}fringence Magn\'{e}tique du Vide (BMV) experiment one of the coefficients, $\gamma\approx 10^{-20}$ T$^{-2}$, was estimated. The canonical, symmetrical Belinfante energy-momentum tensors and dilatation current were obtained. The dilatation symmetry and the dual symmetry are broken in the model considered.

\end{abstract}

\section{Introduction}

After results of the PVLAS Collaboration \cite{PVLAS} nonlinear electrodynamics is of great interest.
Experimental discovery of the effect of birefringence could indicate new physics. Linear Maxwell's electrodynamics and the
Born-Infeld (BI) electrodynamics \cite{Born} do not produce the birefringence phenomenon. But QED with quantum
corrections to classical electrodynamics gives the Heisenberg-Euler Lagrangian density \cite{Heisenberg}, \cite{Schwinger} that results in the small effect of birefringence. The similar effect occurs in the model  \cite{Kruglov}, in generalized BI electrodynamics \cite{Kruglov3}, in electrodynamics with Lorentz violation \cite{Kruglov4} and in logarithmic electrodynamics \cite{Gaete}. The exponential nonlinear electrodynamics was considered in \cite{Hendi}. In addition, BI electrodynamics and models \cite{Gaete}, \cite{Kruglov5} give a finite electromagnetic energy of a point charge contrarily to Maxwell's electrodynamics. The dimensional parameter introduced in nonlinear electrodynamics is connected with the upper bound on the possible electric field.

In this paper we generalize the model \cite{Kruglov5} to have the effect of birefringence.
We use the Heaviside-Lorentz system with $\hbar =c=\varepsilon_0=\mu_0=1$ and the metric tensor $g_{\mu\nu}=$diag$(-1,1,1,1)$. Greek letters run from $1$ to $4$ and Latin letters range from $1$ to $3$.

\section{The model}

We suggest the Lagrangian density of nonlinear electrodynamics
\begin{equation}
{\cal L} = -{\cal F}-\frac{a{\cal F}}{2(\beta{\cal F})+1}+\frac{\gamma}{2}{\cal G}^2,
 \label{1}
\end{equation}
where $a$ is a dimensionless parameter and $\beta$, $\gamma$ are parameters with the dimensions of (length)$^4$  ($\beta{\cal F}$ and $\gamma {\cal G}$ are dimensionless). The motivation for Eq. (1) is based on the fact that the electric displacement field in this model is singular at $\textbf{B}=0$, $\textbf{E}=1/\sqrt{\beta}$ (see Eq. (6)), and this allows us to find the finite electric field at the origin of a point-like charged particle \cite{Kruglov5}. The same feature takes place in BI electrodynamics. The second quadratic term in Eq. (1) is similar to weak-field  Heisenberg-Euler limit that gives the contribution to birefringence. One can consider a more general dependence on ${\cal G}$ in Eq. (1) that will change the indices of refraction but this generalization does not contribute to electrostatics and maximum possible electric field.
The Lorentz-invariants are ${\cal F}=(1/4)F_{\mu\nu}F^{\mu\nu}=(\textbf{B}^2-\textbf{E}^2)/2$, ${\cal G}=(1/4)F_{\mu\nu}\tilde{F}^{\mu\nu}=\textbf{E}\cdot \textbf{B}$, $\tilde{F}_{\mu\nu}=(1/2)\varepsilon_{\mu\nu\alpha\beta}F^{\alpha\beta}$ is a dual tensor, and $F_{\mu\nu}=\partial_\mu A_\nu-\partial_\nu A_\mu$ is the field strength ($A_\mu$ is the 4-vector-potential). At $\gamma=0$ we arrive at the model introduced in \cite{Kruglov5}.
It was shown that the parameter $\beta$ gives the maximum of the field strength in the origin of the point-like electric charge. The electrostatic energy of the point-like electric charge is also finite in the model \cite{Kruglov5}. The same phenomenon occurs in the model based on the Lagrangian density (1) because the last term in Eq.(1) does not contribute to electrostatics ($\textbf{B}=0$).
At $a\rightarrow 0$, $\gamma=0$ the Lagrangian density (1) becomes the Maxwell Lagrangian density. If $\beta{\cal F}\ll 1$, one finds from Eq. (1) the approximate Lagrangian density
\begin{equation}
{\cal L}\approx -(1+a){\cal F}+2a\beta {\cal F}^2+\frac{\gamma}{2}{\cal G}^2.
\label{2}
\end{equation}
It is natural to imply that $a\ll 1$. To have the standard linear term in Eq. (2) one can renormalize the fields so that $\textbf{B}'=\sqrt{1+a} \textbf{B}$, $\textbf{E}'=\sqrt{1+a} \textbf{E}$. Then the value $(1+a)$ be absorbed in the first term in Eq. (2) and the Lagrangian density (2) becomes
\begin{equation}
{\cal L}\approx  -{\cal F}'+\frac{2a\beta {\cal F'}^2}{1+a}+\frac{\gamma {\cal G'}^2}{2(1+a)}\approx
-{\cal F}'+2a\beta {\cal F'}^2+\frac{\gamma (1-a){\cal G'}^2}{2}
\label{3}
\end{equation}
in the order of ${\cal O}(a^2)$.
We will use the exact Lagrangian (1) without the approximation leading to Eqs. (2),(3). The model based on the Lagrangian density (3), was considered in \cite{Kruglov}. That model admits the phenomena of vacuum birefringence when the external constant magnetic field is present. Thus, the phase velocities of light are different for
different polarizations of electromagnetic fields. We will show that vacuum birefringence also takes place in the model based on Eq. (1). At
\begin{equation}
a\beta=\frac{4\alpha^2}{45m_e^4},~~~~\gamma (1-a)=\frac{28\alpha^2}{45m_e^4},
\label{4}
\end{equation}
in Eq. (3), where $\alpha=e^2/(4\pi)=1/137$ is the fine structure constant, $m_e$ is the electron mass, we arrive at
the weak-field ($\alpha {\cal F}/m_e^4\ll 1$) Heisenberg-Euler limit \cite{Heisenberg}, \cite{Schwinger} due to
one-loop quantum corrections in QED. The first three orders in the weak-field expansion are
presented, for example, in \cite{A}.
The nonlinear model of electrodynamics based on Eq. (3) naturally appears due to the vacuum polarization of arbitrary spin particles \cite{Kruglov}. The effective models of electromagnetic fields that lead to birefringence are also considered in \cite{Kruglov3}, \cite{Kruglov4}.

From Eq. (1) with the help of the Euler-Lagrange equations we obtain the equations of motion
\begin{equation}
\partial_\mu\left(F^{\mu\nu}+\frac{aF^{\mu\nu}}{\left[2(\beta{\cal F})+1\right]^2}-\gamma {\cal G}\tilde{F}^{\mu\nu}\right)=0.
\label{5}
\end{equation}
From Eq. (1) we find the electric displacement field
$\textbf{D}=\partial{\cal L}/\partial \textbf{E}$:
\begin{equation}
\textbf{D}=\left(1+\frac{a}{\left[\beta \left(\textbf{B}^2-\textbf{E}^2\right)+1\right]^2}\right)\textbf{E}+\gamma\left(\textbf{E}\cdot\textbf{B}\right)\textbf{B}.
\label{6}
\end{equation}
Eq. (6) can be represented in the tensor form $D_i=\varepsilon_{ij}E_j$ where the electric permittivity $\varepsilon_{ij}$ is
\begin{equation}
\varepsilon_{ij}=\varepsilon\delta_{ij}+\gamma B_iB_j,~~~~\varepsilon=1+\frac{a}{\left[\beta \left(\textbf{B}^2-\textbf{E}^2\right)+1\right]^2}.
\label{7}
\end{equation}
We obtain the magnetic field $\textbf{H}=-\partial{\cal L}/\partial \textbf{B}$:
\begin{equation}
\textbf{H}= \left(1+\frac{a}{\left[\beta \left(\textbf{B}^2-\textbf{E}^2\right)+1\right]^2}\right)\textbf{B}-\gamma\left(\textbf{E}\cdot\textbf{B}\right)\textbf{E}.
\label{8}
\end{equation}
Eq. (8) may be written as $\textbf{B}_i=\mu_{ij}\textbf{H}_j$, where the inverse magnetic permeability $(\mu^{-1})_{ij}$ is given by
\begin{equation}
(\mu^{-1})_{ij}=\varepsilon\delta_{ij}-\gamma E_iE_j.
\label{9}
\end{equation}
From Eq. (9) we obtain the magnetic permeability tensor
\[
\mu_{ij}=\varepsilon^{-1}\delta_{ij}-\frac{\gamma}{\varepsilon\left(\gamma E^2-\varepsilon\right)} E_iE_j.
\]
Eq.(5) is equivalent to the first pair of the Maxwell equations
\begin{equation}
\nabla\cdot \textbf{D}= 0,~~~~ \frac{\partial\textbf{D}}{\partial
t}-\nabla\times\textbf{H}=0.
\label{10}
\end{equation}
The Bianchi identity $\partial_\mu \widetilde{F}^{\mu\nu}=0$ gives the second pair of Maxwell's equations
\begin{equation}
\nabla\cdot \textbf{B}= 0,~~~~ \frac{\partial\textbf{B}}{\partial
t}+\nabla\times\textbf{E}=0.
\label{11}
\end{equation}
As the electric permittivity tensor $\varepsilon_{ij}$ and tensor of magnetic permeability $\mu_{ij}$ depend on the electromagnetic fields, equations (10),(11) represent the nonlinear Maxwell equations.
The vacuum of the model mimics a medium with complicated properties. The speed of propagation of
waves depends on polarization.
We will investigate this phenomenon in the next section.

From Eqs. (6), (8) we obtain $\textbf{D}\cdot\textbf{H}=\textbf{E}\cdot\textbf{B}\left(\varepsilon^2+2\varepsilon\gamma{\cal F}-\gamma^2{\cal G}^2\right)$. Therefore $\textbf{D}\cdot\textbf{H}\neq\textbf{E}\cdot\textbf{B}$ and according to \cite{Gibbons}
the dual symmetry is violated.

\section{Vacuum magnetic birefringence}

Vacuum birefringence within QED was investigated in \cite{Dittrich}, \cite{Adler1}, \cite{Biswas}, \cite{Battesti}.
We consider now the plane electromagnetic waves $(\textbf{e}, \textbf{b}$) propagating in $z$-direction at the presence of a constant and uniform magnetic induction field $\textbf{B}_0=(0,B_0,0)$ where
\begin{equation}
\textbf{e}=\textbf{e}_0\exp\left[-i\left(\omega t-kz\right)\right],~~~\textbf{b}=\textbf{b}_0\exp\left[-i\left(\omega t-kz\right)\right].
\label{12}
\end{equation}
The resultant electromagnetic fields are given by $\textbf{E}=\textbf{e}$, $\textbf{B}=\textbf{b}+\textbf{B}_0$. We assume that electromagnetic wave fields $\textbf{e}, \textbf{b}$ are weaker compared with the strong constant and uniform magnetic induction field, i.e. $e_0,b_0\ll B_0$. Then after linearizing Eqs. (7),(9) we obtain the electric permittivity and magnetic permeability
\begin{equation}
\varepsilon_{ij}=\varepsilon\delta_{ij}+\gamma \delta_{i2}\delta_{j2}B_{0}^2,~~~\varepsilon=1+\frac{a}{\left(\beta B_0^2+1\right)^2},~~~\mu_{ij}=\mu\delta_{ij},~~~\mu=\varepsilon^{-1}.
\label{13}
\end{equation}
For these constants, $\varepsilon_{ij}$ and $\mu$, we get from Maxwell's equations (10), (11) the wave equation
\begin{equation}
\mu\varepsilon_{ij}\partial^2_tE_j-\triangle E_i+\partial_i\partial_jE_j=0,
\label{14}
\end{equation}
where $\triangle=\nabla^2$. Let us consider different polarizations of electromagnetic waves. For the case when polarization is parallel to external magnetic field, $\textbf{e}=e(0,1,0)$, we find from Eq. (14) $\mu\varepsilon_{22}\omega^2=k^2$. As a result, one obtains the index of refraction
\begin{equation}
n_\|=\sqrt{\mu\varepsilon_{22}}=\sqrt{1+\frac{\gamma B_0^2\left(\beta B_0^2+1\right)^2}{a+\left(\beta B_0^2+1\right)^2}}.
\label{15}
\end{equation}
If polarization (the direction of the electric field) is perpendicular to external magnetic field, $\textbf{e}=e(1,0,0)$, we obtain $\mu\varepsilon\omega^2=k^2$, and the index of refraction is
\begin{equation}
n_\perp=\sqrt{\varepsilon\mu}=1.
\label{16}
\end{equation}
In this case the magnetic field of the plane wave is parallel to external constant and uniform magnetic field and,
as a result, the refractive index is exactly unity.
Thus, we arrive at the effect of vacuum birefringence: phase velocity depends on polarization. When polarization of the electromagnetic wave is parallel to external magnetic field the velocity reduces, $v=c/n_\|$, and becomes less than the speed of light $c=1$. For the case $\textbf{e}\perp \textbf{B}_0$ the speed of the electromagnetic wave is $c=1$.

A linear birefringence in the presence of a transverse external magnetic field is called the Cotton-Mouton (CM) effect \cite{Battesti}. The difference in indices of refraction is defined as
\begin{equation}
\triangle n_{CM}=n_\|-n_\perp=k_{CM}\textbf{B}_0^2.
\label{17}
\end{equation}
From Eqs. (15),(16) we obtain, using approximation $\beta \textbf{B}_0^2\ll 1$, $\gamma \textbf{B}_0^2\ll 1$,
\begin{equation}
\triangle n_{CM}=\sqrt{1+\frac{\gamma B_0^2\left(\beta B_0^2+1\right)^2}{a+\left(\beta B_0^2+1\right)^2}}-1
\approx \frac{\gamma (1-a)}{2}\textbf{B}_0^2,
\label{18}
\end{equation}
and CM coefficient becomes
\begin{equation}
k_{CM}\approx \frac{\gamma (1-a)}{2}.
\label{19}
\end{equation}
It is interesting that at $\gamma =0$ the effect of birefringence vanishes. This conclusion is based on calculations using the exact Lagrangian (1). If one explores the approximate Lagrangian (3), motivated by QED, the term containing the parameter $\beta$ gives the contribution to birefringence. Therefore the model under consideration is not trivial and allows us to obtain bounds on the parameter $\gamma$ from magnetic birefringence experiment.
The measurement of the vacuum magnetic linear birefringence by the BMV experiment for a maximum field of $B_0=6.5$ T
gave \cite{Rizzo}
\begin{equation}
k_{CM}=(5.1\pm 6.2)\times 10^{-21} \mbox {T}^{-2}.
\label{20}
\end{equation}
From Eqs. (19),(20) at $a\ll 1$, we obtain the approximate value
\begin{equation}
\gamma \approx 10^{-20} \mbox {T}^{-2}.
\label{21}
\end{equation}
It should be mentioned that the value predicted by QED, using one loop approximation, is \cite{Rizzo} $k_{CM}\approx 4.0\times 10^{-24} \mbox {T}^{-2}$. Therefore, new experiments with higher accuracy are needed to verify the QED prediction.

\section{The energy-momentum tensor and dilatation current}

With the help of Eq. (1) and the expression of the canonical energy-momentum tensor we find
\begin{equation}
T_{\mu\nu}^{(c)}=-(\partial_\nu A^\alpha)\left(F_{\mu\alpha}\varepsilon-\gamma{\cal G}\tilde{F}_{\mu\alpha}\right)-g_{\mu\nu}{\cal L},
\label{22}
\end{equation}
so that $\partial_\mu T^{(c)\mu\nu}=0$. To obtain the symmetric Belinfante tensor we use the relation \cite{Coleman}:
\begin{equation}
T_{\mu\nu}^{(B)}=T_{\mu\nu}^{(c)}+\partial_\beta X^{\beta}_{~\mu\nu},
\label{23}
\end{equation}
where
\begin{equation}
X_{\beta\mu\nu}=\frac{1}{2}\left[\Pi_{\beta}^{~\sigma}\left(\Sigma_{\mu\nu}\right)_{\sigma\rho}
-\Pi_{\mu}^{~\sigma}\left(\Sigma_{\beta\nu}\right)_{\sigma\rho}-
\Pi_{\nu}^{~\sigma}\left(\Sigma_{\beta\mu}\right)_{\sigma\rho}\right]A^\rho,
\label{24}
\end{equation}
\begin{equation}
\Pi^{\mu\sigma}=\frac{\partial{\cal L}}{\partial(\partial_\mu
A_\sigma)}=-F^{\mu\sigma}\varepsilon+\gamma{\cal G}\tilde{F}^{\mu\sigma},
\label{25}
\end{equation}
and $X_{\beta\mu\nu}=-X_{\mu\beta\nu}$. Because $\partial_\mu\partial_\beta X^{\beta\mu\nu}=0$, one has conserved Belinfante tensor, $\partial_\mu T^{(B)\mu\nu}=\partial_\mu T^{(c)\mu\nu}=0$. The matrix elements of the generators of the Lorentz transformations $\Sigma_{\mu\alpha}$ are given by $\left(\Sigma_{\mu\alpha}\right)_{\sigma\rho}=g_{\mu\sigma}g_{\alpha\rho}
-g_{\alpha\sigma}g_{\mu\rho}.$
With the help of Eq. (24) we find $\partial_\beta X^{\beta\mu\nu}=\Pi^{\beta\mu}\partial_\beta A^\nu.$
From equation of motion (5), $\partial_\mu\Pi^{\mu\nu}=0$, and
Eqs. (24),(25), one obtains the Belinfante tensor (23)
\begin{equation}
T_{\mu\nu}^{(B)}=-F_{\nu}^{~\alpha}\left(F_{\mu\alpha}\varepsilon-\gamma{\cal G}\tilde{F}_{\mu\alpha}\right)-g_{\mu\nu}{\cal L}.
\label{26}
\end{equation}
The trace of the energy-momentum tensor (26) is given by
\begin{equation}
T_{~\mu}^{(B)\mu}=\frac{8a\beta {\cal F}^2}{\left[2(\beta{\cal F})+1\right]^2}+2\gamma {\cal G}^2.
\label{27}
\end{equation}
Putting $a=0$, $\gamma=0$ we arrive at linear electrodynamics and the trace of the energy-momentum tensor (27) vanishes.
According to \cite{Coleman}, we have the modified dilatation current $D_{\mu}=x^\alpha T_{\mu\alpha}^{(B)}+V_\mu,$
with the field-virial $
V_\mu=\Pi^{\alpha\beta}\left[g_{\alpha\mu}g_{\beta\rho}
-\left(\Sigma_{\alpha\mu}\right)_{\beta\rho}\right]A^\rho=0.$
Therefore, the modified dilatation current is given by
\begin{equation}
D_{\mu}=x^\alpha T_{\mu\alpha}^{(B)}.
\label{28}
\end{equation}
Then the 4-divergence of dilatation current is
\begin{equation}
\partial_\mu D^{\mu}=T_{~\mu}^{(B)\mu}.
\label{29}
\end{equation}
As a result, the dilatation symmetry is broken because of the presence of the
dimensional parameters $\beta$ and $\gamma$. The same situation occurs in BI electrodynamics \cite{Kruglov3}.
The conformal symmetry is also broken \cite{Coleman} because the dilatation symmetry is a subgroup of the conformal group.

\section{Conclusion}

A new model of nonlinear electrodynamics with three parameters $a$,$\beta$ and $\gamma$ is suggested.  The variables $\beta^{1/4}$, $\gamma^{1/4}$ have dimensions of the length and can appear, probably, due to some gravity
effects. The canonical and symmetrical Belinfante energy-momentum tensors and dilatation current were obtained showing that the dilatation symmetry is broken in the model. The scale symmetry is violated because the dimensional parameters $\beta$, $\gamma$ were introduced.

We have obtained the phenomenon of vacuum birefringence if the external constant and uniform induction magnetic field is present. The indices of refraction for two polarizations of electromagnetic waves, parallel and perpendicular to the magnetic field were calculated. From the BMV experiment the coefficient $\gamma\approx 10^{-20}$ T$^{-2}$ was estimated. The model suggested is not trivial because in vacuum birefringence only the parameter $\gamma$ enters and new kind of experiments are needed to obtain bounds on parameters $a$ and $\beta$. In addition, this model gives the finite electric energy of point-like charged particle similar to BI electrodynamics. The BMV experiment does not confirm the QED prediction of the value $k_{CM}$, and new experiments, with higher orders of magnitude, are needed. Possibly, to explain the experimental data, we have to modify the linear Maxwell equations to take into account some corrections going from new physics.

\end{document}